\documentclass[twoside]{article}
\usepackage{graphicx} 
\usepackage{aistats2024}
\usepackage{multirow}
\usepackage[ruled,vlined]{algorithm2e}
\usepackage{lipsum} 

\begin{document}

%

%

\twocolumn[

\aistatstitle{Medical Image Segmentation via Sparse Coding Decoder}
\def\statePaper{1}
\aistatsauthor{ Long Zeng  \And Kaigui Wu }

\aistatsaddress{ 
Chongqing University 
 \And 
Chongqing University } ]

\begin{abstract}
 Transformers have achieved significant success in medical image segmentation, owing to its  capability to capture long-range dependencies. Previous works incorporate convolutional layers into the encoder module of transformers, thereby enhancing their ability to learn local relationships among pixels. However, transformers may suffer from limited generalization capabilities and reduced robustness, attributed to the insufficient spatial recovery ability of their decoders. To address this issue, A convolution sparse vector coding based decoder is proposed , namely CAScaded multi-layer Convolutional Sparse vector Coding DEcoder (CASCSCDE), which represents features extracted by the encoder using sparse vectors. To prove the effectiveness of our CASCSCDE, The widely-used TransUNet model is chosen for the demonstration purpose, and the CASCSCDE is incorporated with TransUNet to establish the TransCASCSCDE architecture. Our experiments demonstrate that TransUNet with CASCSCDE significantly enhances performance on the Synapse benchmark, obtaining up to 3.15\% and 1.16\% improvements in DICE and mIoU scores, respectively. CASCSCDE opens new ways for constructing decoders based on convolutional sparse vector coding.

\end{abstract}

\begin{figure*}[!t] 
\centering 
\vspace{.3in}
\includegraphics[width=\textwidth]{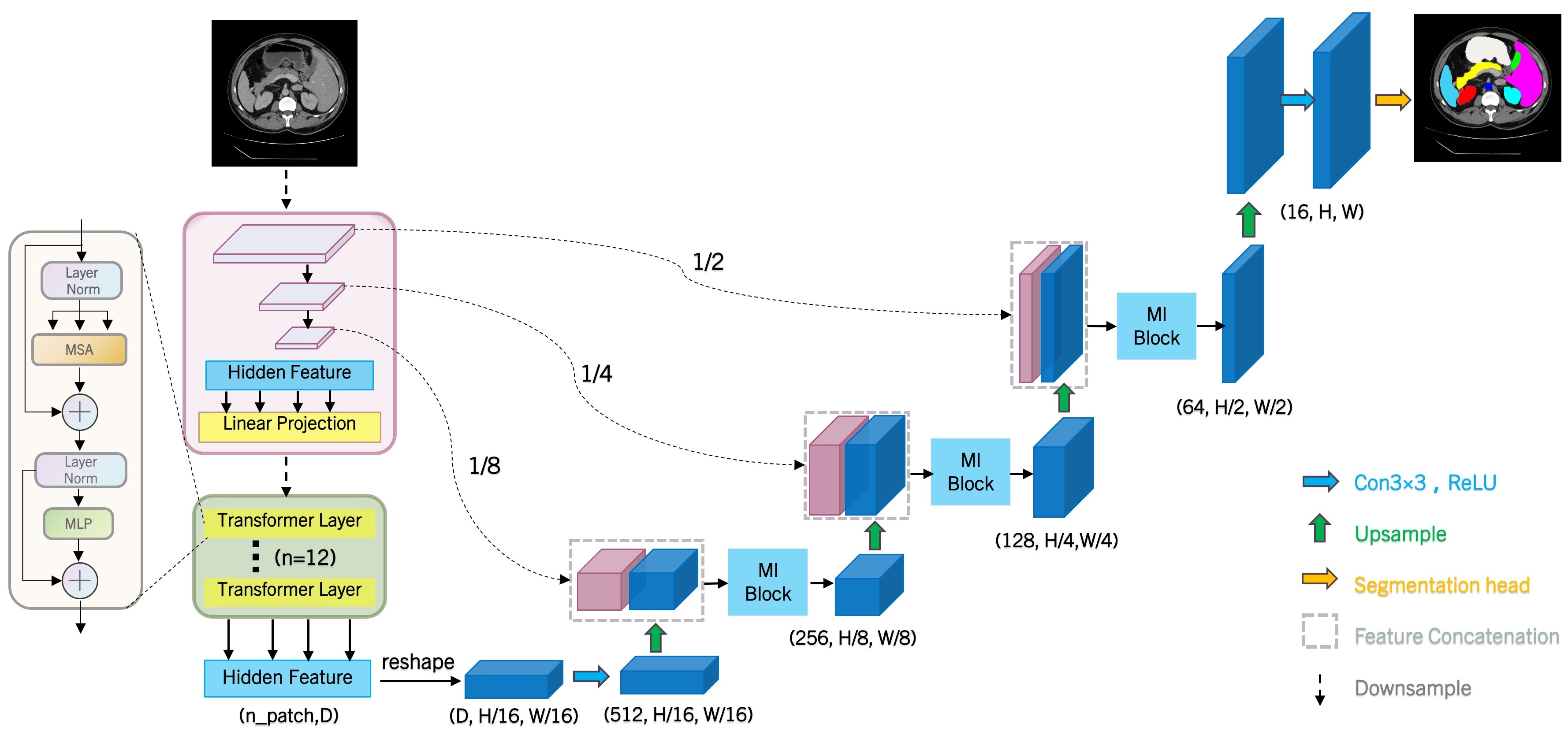} 
\vspace{.3in}
\caption{Overview of the TransCASCSCDE.}
\vspace{2\baselineskip} 
\end{figure*}

\begin{algorithm*}
\caption{ML-block Neural Network Module}
\KwData{Input tensor x, T}
\KwResult{Output tensor gamma2}
\Begin{
    Initialize  W1, W2, c1, c2, b1, b2 \;   
    gamma1 = ReLU(BN(c1 * Conv2D(x, W1, stride, padding) + b1))\;
    gamma2 = ReLU(BN(c2 * Conv2D(gamma1, W2, stride, padding) + b2))\;
    \For{t = 1 \KwTo T}{
        gamma1 = ConvTranspose2D(gamma2, W2, stride, padding)\;
        temp1 = ConvTranspose2D(gamma1, W1, stride, padding) - x\;
        temp2 = Conv2D(temp1, W1, stride=1, padding)\;
        gamma1 = ReLU(gamma1 - c1 * temp2 + b1)\;
        temp3 = ConvTranspose2D(gamma2, W2, stride, padding) - gamma1\;
        temp4 = Conv2D(temp3, W2, stride, padding)\;
        gamma2 = ReLU(gamma2 - c2 * temp4 + b2)\;
    }
    \Return{gamma2}\;
}
\end{algorithm*}

\section{Introduction}

In the realm of disease diagnosis, the significance of automated medical image segmentation has grown exponentially. The advent of U-shaped convolutional neural networks (CNNs), spearheaded by the development of UNet (Ronneberger et al., 2015), has set a new standard in the field. Variations of UNet, notably UNet++ (Zhou et al., 2018) and UNet3Plus (Huang et al., 2020), have further solidified this paradigm by generating detailed segmentation maps, leveraging multi-stage features through skip connections.

However, a persistent challenge with CNN-based techniques is their constrained capability to discern long-range pixel relationships (Cao et al., 2021). Some works (Chen et al., 2018; Oktay et al., 2018; Fan et al., 2020) try to address this issue by incorporating attention mechanisms within encoders or decoders. Despite these significant efforts, the CNN-based methods remain limited in capturing long-range dependencies. As Vision Transformers (Dosovitskiy et al., 2020) have come to the forefront, many works (Cao et al., 2021; Chen et al., 2021; Dong et al., 2021; Wang et al., 2022b) have explored transformer encoders to address the above problem, particularly in medical image segmentation.

Transformers utilize self-attention (SA) to discern long-range dependencies by understanding correlations across input patches. The recent introduction of hierarchical vision transformers, including the pyramid vision transformer (PVT) (Wang et al., 2021) with its spatial reduction focus, the Swin transformer (Liu et al., 2021) emphasizing window-based attention, and MaxViT (Tu et al., 2022) featuring multi-axis attention, has enhanced performance. Notably, these advanced vision transformers have proven highly effective in medical image segmentation (Cao et al., 2021; Dong et al., 2021; Wang et al., 2022b).

The encoders of these transformer-based architectures possess strong information extraction capabilities. However, their decoders lack the ability to recover spatial detail information, and they have limited generalization and robustness. To address this limitation, we introduce a CAScaded multi-layer Convolutional Sparse vector Coding DEcoder (CASCSCDE) which effectively extracts important features and ignores irrelevant information, thereby improving the accuracy of segmentation. CASCSCDE employs multi-layer convolutional operations to achieve sparse vector encoding. Due to its sparse encoding characteristics, it can identify and ignore noise in images, thus achieving better segmentation results even in medical images with noise. It is adaptable to various types of medical images and tasks, demonstrating good versatility. Our contributions are summarized as follows:
\begin{itemize}
    \item \textbf{Novel Network Architecture:} A cascaded multi-layer convolution sparse vector coding decoder (CASCSCDE) is introduced for 2D medical image segmentation, which takes advantage of the sparsity and robustness brought by sparse coding. We build our decoder using a multi-layer convolutional module, which captures more complex semantic information and suppresses unnecessary information. To the best of our knowledge, we are the pioneers in introducing this kind of decoder for medical image segmentation
\end{itemize}

\begin{itemize}
\item \textbf{Significant Performance Enhancement}: It is demonstrated that CASCSCDE can be used in conjunction with hierarchical vision encoders (e.g., TransUNet (Chen et al.,2021)), significantly enhancing the performance of 2D medical image segmentation. Compared to the original baseline, transCASCSCDE achieves significantly improved performance on the Synapse multi-organ segmentation benchmark. 
\end{itemize}

\section{Related Works}

\subsection{Sparse Vector Encoding in Image Segmentation}

Sparse representations, offering a compact and interpretable way to represent signals, have been a cornerstone in image processing. The domain was pioneered by Chen et al., who introduced atomic decomposition using basis pursuit, a technique that identifies the optimal basis from a vast dictionary to represent a signal (Chen et al., 2001). This foundational approach was further elaborated by Elad, who provided an exhaustive overview of both the theoretical and practical aspects of sparse and redundant representations in signal and image processing (Elad, 2010).

The intersection of sparse coding with deep learning, especially convolutional neural networks (CNNs), has garnered significant attention in recent years. Papyan et al. (Papyan et al., 2017) made a groundbreaking contribution by analyzing CNNs through the perspective of convolutional sparse coding. This analysis not only offered insights into the interpretability of CNNs using sparse coding techniques but also paved the way for the integration of these techniques into deep learning models. Building on this, Papyan et al. (Papyan et al., 2018) introduced a multi-layer sparse model, emphasizing its intrinsic connection to CNNs and shedding light on the intricate workings of deep neural architectures.

In the realm of image segmentation, the fusion of sparse representations and deep learning has shown immense promise. The introduction of the multi-layer convolutional sparse coding block (Papyan et al., 2017) has been a game-changer. This block refines feature maps in segmentation models, leading to faster convergence and a more nuanced extraction of image semantics and appearance nuances. Sulam et al. (Sulam et al., 2019) further explored this synergy, emphasizing the potential of sparse representations in bolstering the performance of convolutional networks, particularly for image segmentation tasks. Due to its potential, we incorporate the sparse vector encoding with our decoder.
\subsection{Medical image segmentation}
Medical image segmentation stands as a pivotal research area in the fields of computer vision and medical imaging. Among the techniques employed, the U-shaped encoder-decoder architecture (Ronneberger et al., 2015) has gained prominence. UNet efficiently aggregates features from the encoder with upsampled features from the decoder through skip connections, yielding high-resolution segmentation maps. Variants like UNet++ (Zhou et al., 2018) employ nested and dense skip connections to enhance feature propagation and bridge the semantic gap, while UNet 3+ (Huang et al., 2020) incorporates full-scale skip connections, including intra-connections among decoder blocks, to achieve refined segmentation outcomes. CSC-Unet (Tang et al., 2021) employs a sparse coding strategy to construct a U-shaped encoder-decoder architecture, offering a approach to segmentation tasks.

Currently, transformers exhibit outstanding performance in the realm of medical image segmentation. TransUNet (Chen et al., 2021) stands out by integrating a hybrid CNN-transformer encoder to capture long-range dependencies, further complemented by a cascaded CNN upsampler as its decoder to pinpoint local contextual relations. In a similar vein, Swin-Unet (Cao et al., 2021), a pure transformer model, is constructed upon the foundational Swin transformer, harnessing transformers in both its encoder and decoder stages for optimized segmentation. Due to the remarkable performance of these models, The hybrid CNN-transformer encoder is integrated with the CASCSCDE to form a unified architecture.

\begin{table*}[htbp]
  \centering
  \caption{Results of Synapse multi-organ segmentation. Only DICE scores are reported for individual organs. R50+UNet(RUNet) and R50+AttnUNet(RAUNet) adopt a pre-trained ResNet50 backbone network.The results of UNet, AttnUNet, SSFormerPVT(SPVT), PolypPVT(PPVT), and TransUNet are taken from  MERIT (Rahman et al., 2023). The results of TFCNs and SwinUNet are taken from PVT-CASCADE (Rahman et al., 2023).  ↑ denotes higher the better, ↓ denotes lower the better. }\label{sample-table}

  \fontsize{8.5pt}{9.5pt}\selectfont
  \setlength{\tabcolsep}{5.5pt}
  \renewcommand{\arraystretch}{1.5}
  \begin{tabular}{l c c c c c c c c c c}
    \\
    \multirow{2}{*}{\textbf{Framework}} & \multicolumn{2}{c}{\textbf{Average}}  & \multirow{2}{*}{\textbf{AT}} & \multirow{2}{*}{\textbf{GB}} & \multirow{2}{*}{\textbf{KL}} & \multirow{2}{*}{\textbf{KR}} & \multirow{2}{*}{\textbf{Liver}} & \multirow{2}{*}{\textbf{PC}} & \multirow{2}{*}{\textbf{SP}} & \multirow{2}{*}{\textbf{SM}} \\[-0.3cm]
    &  \multirow{2}{*}{\textbf{DSC} $\uparrow$}& \multirow{2}{*}{\textbf{HD} $\downarrow$} & & & & & & &\\[0.2cm]
    
    \hline 
    UNet (Ronneberger et al.,2015) & 70.11 & 44.69 & 84.00 & 56.70 & 72.41 & 62.64 & 86.98 & 48.73 & 81.48 & 67.96 \\
    AttnUNet (Oktay et al.,2018) & 71.70 & 34.47 & 82.61 & 61.94 & 76.07 & 70.42 & 87.54 & 46.70 & 80.67 & 67.66 \\
    R50+UNet (Chen et al.,2021)& 74.68 & 36.87 & 84.18 & 62.84 & 79.19 & 71.29 & 93.35 & 48.23 & 84.41 & 73.92 \\
    R50+AttnUNet (Chen et al.,2021) & 75.57 & 36.97 & 55.92 & 63.91 & 79.20 & 72.71 & 93.56 & 49.37 & 87.19 & 74.95 \\
    SSFormerPVT (Wang et al.,2022) & 78.01 & 25.72 & 82.78 & 63.74 & 80.72 & 78.11 & 93.53 & \textbf{61.53} & 87.07 & 76.61 \\
    PolypPVT (Dong et al.,2021) & 78.08 & 25.61 & 82.34 & 66.14 & 81.21 & 73.78 & 94.37 & 59.34 & 88.05 & 79.4 \\
    TFCNs(Li et al,.2022) & 75.63 & 30.63 & 88.23 & 59.18 & 80.99 & 73.12 & 92.02 & 54.24 & 88.36 & 68.9 \\
    TransUNet (Chen et al.,2021) & 77.48 & 31.69 & 87.23 & 63.13 & 81.87 & 77.02 & 94.08 & 55.86 & 85.08 & 75.62 \\
    SwinUNet (Cao et al., 2021) & 77.58 & \textbf{27.32} & 81.76 & 65.95 & 82.32 & \textbf{79.22} & 93.73 & 53.81 & 88.04 & 75.79 \\
    \hline 
    TransCASCSCDE (ours) & \textbf{80.63} & 28.53 & \textbf{88.35} & \textbf{67.19} & \textbf{84.00} & 78.13 & \textbf{96.02} & 60.53 & \textbf{88.55} & \textbf{82.31} \\
    \hline 
    Improve TransUNet & 3.15 &  1.16 & 1.12 & 4.06 & 2.13 & 1.11 & 1.94 & 4.67 & 3.47 & 6.69 \\
    
    \hline \\
  \end{tabular}
  
\end{table*}

\section{Method}
This section, First introduces the transformer backbones and CASCSCDE, and then describe a transformer-based architecture(TransCASCSCDE) incorporating our proposed decoder.

\subsection{Transformer backbones}
We use the hybrid CNN-transformer (instead of only CNN) as the encoder to ensure enough generalization and multi-scale feature processing abilities for medical image segmentation. TransUNet (Chen et al., 2021) utilizes a transformer on top of CNN to capture both global and spatial relationships among features. Our proposed decoder is flexible and easy to adopt with other hierarchical backbone networks. It is also suitable for use in other backbone networks.

\subsection{CAScaded multi-layer Convolutional Sparse vector Coding DEcoder (CASCSCDE)}
Existing transformer-based models face challenges in removing noise and redundant information, as well as accurately reconstructing the boundaries of target areas when capturing key features in images. To address this issue, we propose a Multi-Layer decoder, CASCSCDE.

As shown in Figure 1, CASCSCDE consists of the Up Conv block to upsample the features, the skip connection serves to both preserve spatial resolution and enrich contextual information, and the ML (Multi-Layer) block for Sparse Coding. The decoder starts by upsampling the compressed feature maps received from the transformer layers. After upsampling, the decoder concatenates additional feature maps from the encoder through skip connections to the upsampled feature maps. Then, we process the concatenated features using our ML-block to further refine the features.

\subsubsection{Multi Layer Convolution Sparse Coding Block(ML-Block)}

As illustrated in Algorithm 1, the ML-block operates through a sequence of steps as detailed in the previously provided pseudocode. The core procedures encompass encoding, during which gamma1 and gamma2 are computed, succeeded by a loop executing backward computation for T iterations. A larger value of T results in sparser feature vectors. This block is meticulously crafted to execute sparse vector encoding within the decoder segment of the network. The primary role of the ML-block is to iteratively sparsify feature vectors through multi-layer convolution. By adopting this approach, the network is better equipped to capture both the semantic and appearance nuances of the image, reproduce spatial details more accurately, and suppress noise. Consequently, this leads to a notable enhancement in segmentation performance.

\section{Experiment}

In this section, The transCASCSCDE model is compared with transUNet and other mainstream methods to demonstrate our superior performance. Subsequently, ablation study is conducted on our CASCSCDE decoder. 

\begin{table*}[htbp]
  \centering
  \caption{Results of Ablation study.}\label{sample-table}

  \footnotesize
  \setlength{\tabcolsep}{5.5pt}
  \renewcommand{\arraystretch}{1.5}
  \begin{tabular}{l c c c c c c c c c c}
    \\
    \multirow{2}{*}{\textbf{Framework}} & \multicolumn{2}{c}{\textbf{Average}}  & \multirow{2}{*}{\textbf{AT}} & \multirow{2}{*}{\textbf{GB}} & \multirow{2}{*}{\textbf{KL}} & \multirow{2}{*}{\textbf{KR}} & \multirow{2}{*}{\textbf{Liver}} & \multirow{2}{*}{\textbf{PC}} & \multirow{2}{*}{\textbf{SP}} & \multirow{2}{*}{\textbf{SM}} \\[-0.3cm]
    &  \multirow{2}{*}{\textbf{DSC} $\uparrow$}& \multirow{2}{*}{\textbf{HD} $\downarrow$} & & & & & & &\\[0.2cm]
    
  \hline 
TransCASCSCDE (T=1) & 77.4 & 36.02 & 87.31 & 63.14 & 80.30 & 72.98 & 94.36 & 57.51 & 84.71 & 79.18 \\

TransCASCSCDE (T=2) & \textbf{80.63} & 28.53 & 88.35 & 67.19 & 84.00 & 78.13 & 96.02 & 60.53 & 88.55 & 82.31 \\

TransCASCSCDE (T=3) & 79.02 & 28.95 & 88.21 & 67.02 & 81.80 & 77.50 & 94.93 & 61.05 & 85.69 & 77.51 \\

TransCASCSCDE (T=4) & 78.05 & 30.52 & 87.17 & 55.70 & 83.51 & 77.87 & 95.09 & 60.69 & 85.93 & 78.45 \\

    \hline \\
  \end{tabular}
  
\end{table*}

\subsection{Datasets and evaluation metrics}
The Synapse multi-organ dataset consists of 30 abdominal CT scans, each containing between 85 and 198 slices with a resolution of 512 × 512 pixels. The dataset encompasses a total of 3779 axial contrast-enhanced abdominal CT images, with voxel spatial resolutions ranging from ([0.54-0.54] × [0.98-0.98] × [2.5-5.0]) mm³. For the purposes of training and validation, the dataset is randomly partitioned into 18 scans, comprising 2212 axial slices for training, and 12 scans for validation. The dataset aims to facilitate the segmentation of eight anatomical structures, including the aorta, gallbladder, left kidney, right kidney, liver, pancreas, spleen, and stomach.

\textbf{Evaluation metrics}: The average Dice Similarity Coefficient (DSC) and average Hausdorff Distance (HD) are used as the evaluation metrics in our experiments on the Synapse Multi-organ dataset.

\subsection{Implementation details}
All our experiments are conducted using Pytorch 1.11.0. Our model is trained on a single NVIDIA RTX 3090Ti GPU with 24GB of memory, leveraging pre-trained weights from ImageNet for the backbone networks. Additionally, The AdamW optimizer is utilized with a learning rate and weight decay of 1e-4.

\textbf{Synapse Multi-organ dataset}. Inheriting from TransUNet (Chen et al., 2021) settings, we set the batch size to 24 and cap our training to 150 epochs for each model. The input resolution is configured at 224×224, with a patch size (P) of 16. For data augmentation, we introduce random flipping and rotation. Our loss function is a combination of cross-entropy and DICE loss.

\subsection{Results}
Our architectures (TransCASCSCDE) is compared with mainstream CNN and transformer-based segmentation methods on the Synapse Multi-organ dataset. 

\subsubsection{Experimental results on Synapse dataset}
The performance of different CNN and transformer-based methods are demonstrated in Table 1. As shown, transformer-based models exhibit superior performance compared to CNN-based models. Our proposed CASCSCDE decoder improves the average DICE and HD95 scores of TransUNet by 3.15\% and 1.16\%, respectively. TransCASCSCDE achieves the best average DICE score (80.63\%) among all other methods. Moreover, TransCASCSCDE demonstrates significant performance improvements in both small and large organ segmentation. For small organs, we observe improvements of 4.05\% in gallbladder, 2.13\% in left kidney, and 1.11\% in right kidney. For large organs, there are improvements of 6.69\% in stomach, 1.94\% in pancreas, and 3.47\% in spleen. This enhanced performance is attributed to CASCSCDE's ability to  multi-layer convolutional architecture coupled with a sparse encoding mechanism. This multi-layer convolutional approach effectively bridges the high-level features extracted by the encoder with the input image's foundational features. Leveraging sparse encoding allows for a finer-grained capture of image nuances, facilitating a more precise boundary reconstruction of the target regions during the decoding process.

\subsection{Ablation study}

\textbf{Effect of the times of  iterations(T)}: Here, we compared the the ability of CASCSCDE decoder to recover spatial detail information of image with different iterations times(T) in ML-block, a larger value of T results in sparser feature vector and the results were shown in the Table 2. These results showed that We can observe that as T increases from 1 to 2, the features become sparser and the model's performance sees a significant improvement. However, as T continues to rise to 3 and 4, the features become even sparser and performance starts to decline.This suggests that more complex image features might require higher T values for sparse vectorization to effectively reconstruct the image. Intricate features carry richer information, and increased T values allow the model to capture this information more effectively. However, setting the T value too high can result in overly sparse feature representations, which could negatively impact the model's performance.

\section{Conclusion}
In this paper, an innovative approach is proposed that employs a cascaded multi-layer convolutional sparse vector coding in decoder which represents the features as sparse vectors during the decoding process. Experimental evidence suggests that CASCADE demonstrates a robust capability for refined boundary reconstruction, adeptly fusing both local and global features. This not only enhances its overall versatility but also bolsters its robustness, as evidenced by a notable improvement of 3.15\% DICE over the baseline TransUNet in Synapse multi-organ segmentation. With the incorporation of multi-layer convolutional sparse coding during the decoding phase, CASCADE exhibits an acute precision in locating organs or lesions—evident from the 1.16 enhancement in the HD95 score. Therefore, our decoder underscores a promising potential in enhancing transformer-driven functionalities in medical domains.

\section*{Checklist}

 \begin{enumerate}

 \item For all models and algorithms presented, check if you include:
 \begin{enumerate}
   \item A clear description of the mathematical setting, assumptions, algorithm, and/or model. [Yes]
   \item An analysis of the properties and complexity (time, space, sample size) of any algorithm. [Yes]
   \item (Optional) Anonymized source code, with specification of all dependencies, including external libraries. [No]
 \end{enumerate}

 \item For any theoretical claim, check if you include:
 \begin{enumerate}
   \item Statements of the full set of assumptions of all theoretical results. [Yes]
   \item Complete proofs of all theoretical results. [Yes]
   \item Clear explanations of any assumptions. [Yes]     
 \end{enumerate}

 \item For all figures and tables that present empirical results, check if you include:
 \begin{enumerate}
   \item The code, data, and instructions needed to reproduce the main experimental results (either in the supplemental material or as a URL). [Yes]
   \item All the training details (e.g., data splits, hyperparameters, how they were chosen). [Yes]
         \item A clear definition of the specific measure or statistics and error bars (e.g., with respect to the random seed after running experiments multiple times). [Yes]
         \item A description of the computing infrastructure used. (e.g., type of GPUs, internal cluster, or cloud provider). [Yes]
 \end{enumerate}

 \item If you are using existing assets (e.g., code, data, models) or curating/releasing new assets, check if you include:
 \begin{enumerate}
   \item Citations of the creator If your work uses existing assets. [Yes]
   \item The license information of the assets, if applicable. [Not Applicable]
   \item New assets either in the supplemental material or as a URL, if applicable. [Not Applicable]
   \item Information about consent from data providers/curators. [Yes]
   \item Discussion of sensible content if applicable, e.g., personally identifiable information or offensive content. [No]
 \end{enumerate}

 \item If you used crowdsourcing or conducted research with human subjects, check if you include:
 \begin{enumerate}
   \item The full text of instructions given to participants and screenshots. [Not Applicable]
   \item Descriptions of potential participant risks, with links to Institutional Review Board (IRB) approvals if applicable. [Not Applicable]
   \item The estimated hourly wage paid to participants and the total amount spent on participant compensation. [Not Applicable]
 \end{enumerate}

 \end{enumerate}

\end{document}